\documentclass[pdftex,superscriptaddress,twocolumn]{revtex4}
\usepackage{multirow,enumerate,epsfig,graphics,amssymb,amsmath,subeqnarray,mathrsfs,color,xcolor}
\usepackage{color,graphics,amssymb,amsmath,subeqnarray,amsthm,subfigure,mathrsfs}

\usepackage[colorlinks=true, citecolor=red, linkcolor=blue ]{hyperref}

\newcommand{\nb}{\mathbf{n}}

\makeatletter
\def\sgn{\mathop{\operator@font sgn}}
\def\threevdots{\vbox{\baselineskip1\p@ \lineskiplimit\z@
  \kern6\p@\hbox{.}\hbox{.}\hbox{.}}}
\makeatother

\begin{document}
\title{Monitoring the orientation of rare-earth-doped
nanorods for flow shear tomography}
\author{Jongwook Kim}
\email{jong-wook.kim@polytechnique.edu}
\affiliation{Laboratoire de Physique de la Matière Condensée, Ecole Polytechnique, CNRS, Université Paris-Saclay, 91128 Palaiseau, France}
\author{S\'ebastien Michelin}
\email{sebastien.michelin@ladhyx.polytechnique.fr}
\affiliation{LadHyX -- D\'epartement de M\'ecanique, Ecole Polytechnique -- CNRS, 91128 Palaiseau, France}
\author{Michiel Hilbers}
\affiliation{van't Hoff Institute for Molecular Sciences, University of Amsterdam, PO Box 94157, 1090 GD Amsterdam, The Netherlands}
\author{Lucio Martinelli}
\affiliation{Laboratoire de Physique de la Matière Condensée, Ecole Polytechnique, CNRS, Université Paris-Saclay, 91128 Palaiseau, France}
\author{Elodie Chaudan}
\affiliation{Laboratoire de Physique de la Matière Condensée, Ecole Polytechnique, CNRS, Université Paris-Saclay, 91128 Palaiseau, France}
\author{Gabriel Amselem}
\affiliation{LadHyX -- D\'epartement de M\'ecanique, Ecole Polytechnique -- CNRS, 91128 Palaiseau, France}
\author{Etienne Fradet}
\affiliation{LadHyX -- D\'epartement de M\'ecanique, Ecole Polytechnique -- CNRS, 91128 Palaiseau, France}
\author{Jean-Pierre Boilot}
\affiliation{Laboratoire de Physique de la Matière Condensée, Ecole Polytechnique, CNRS, Université Paris-Saclay, 91128 Palaiseau, France}
\author{Albert M. Brouwer}
\affiliation{van't Hoff Institute for Molecular Sciences, University of Amsterdam, PO Box 94157, 1090 GD Amsterdam, The Netherlands}
\author{Charles N. Baroud}
\affiliation{LadHyX -- D\'epartement de M\'ecanique, Ecole Polytechnique -- CNRS, 91128 Palaiseau, France}
\author{Jacques Peretti}
\affiliation{Laboratoire de Physique de la Matière Condensée, Ecole Polytechnique, CNRS, Université Paris-Saclay, 91128 Palaiseau, France}
\author{Thierry Gacoin}
\email{thierry.gacoin@polytechnique.edu}
\affiliation{Laboratoire de Physique de la Matière Condensée, Ecole Polytechnique, CNRS, Université Paris-Saclay, 91128 Palaiseau, France}
\date{\today}

\begin{abstract}
Rare-earth phosphors exhibit unique luminescence polarization features originating from the anisotropic symmetry of the emitter ion’s chemical environment. However, to take advantage of this peculiar property, it is necessary to control and measure the ensemble orientation of the host particles with a high degree of precision. Here, we show a methodology to obtain the photoluminescence polarization of Eu-doped LaPO$_4$ nanorods assembled in an electrically modulated liquid- crystalline phase. We measure Eu$^{3+}$ emission spectra for the three main optical configurations ($\sigma$, $\pi$ and $\alpha$, depending on the direction of observation and the polarization axes) and use them as a reference for the nanorod orientation analysis. Based on the fact that flowing nanorods tend to orient along the shear strain profile, we use this orientation analysis to measure the local shear rate in a flowing liquid. The potential of this approach is then demonstrated through tomographic imaging of the shear rate distribution in a microfluidic system.

\end{abstract}
\maketitle

Luminescent particles or molecules are widely used for labelling and tracking of small objects. Anisotropic emitters such as semi-conductor nanowires~\cite{ref1}, quantum rods~\cite{ref2} or organic dyes~\cite{ref3} exhibit polarized luminescence, providing an additional sensitivity to the orientation~\cite{ref4}. The polarization is, in most cases, dominated by the size and shape anisotropy of the emitter particle~\cite{ref1,ref2}, which is understood within the quantum size effect and the electric field confinement effect on both the excitation and emission processes~\cite{ref5,ref6,ref7,ref8}. The photoluminescence of rare-earth phosphors, however, shows a distinguished nature of emission polarization. The photoluminescence spectrum of lanthanide ions in a crystalline host matrix consists of many sharp peaks due to the multiple transition levels within the 4$f$ configuration and their crystal-field splitting into degenerate sub- levels~\cite{ref9,ref10}. Each sublevel emission is polarized along a particular direction allowed by the crystallographic symmetry. Consequently, the emission spectrum from a single crystal manifests variation of its line shape when the crystal’s orientation changes with respect to the direction of polarization analysis~\cite{ref11,ref12,ref13,ref14}. This phenomenon is independent of the particle size and morphology, and is decoupled from the polarization of the usually indirect excitation, which is a crucial advantage for the orientation analysis when compared to the other types of anisotropic emitter.

A prerequisite for precise orientation analysis is to acquire the reference photoluminescence polarization components, which requires either working with a single crystal or achieving a uniform orientation of small crystallites. Here, we use liquid-crystalline (LC) self-assembly of monocrystalline LaPO$_4$:Eu nanorods~\cite{ref15,ref16} that exhibit polarized photoluminescence, as from a large single crystal. By electrically switching the orientation of the LC domain~\cite{ref17}, in a manner similar to the approach by Galyametdinov \emph{et al.} with organic lanthanidomesogens~\cite{ref18}, polarized Eu$^{3+}$ emission spectra could be selectively obtained for the three main configurations ($\sigma$ and $\pi$, the radial propagations polarized perpendicular and parallel to the rod axis, respectively, and $\alpha$, the isotropic axial propagation). We show that the distinct $\sigma$-$\pi$-$\alpha$ line shapes allow us to determine the unknown three-dimensional rod orientation and also the collective degree of orientation of an ensemble of nanorods, thereby establishing a route to the in-situ study of rod-orientation dynamics.

We apply this method to measure the local shear rate in a flowing liquid that imposes the orientation of colloidally dispersed nanorods. The orientation of anisotropic objects under flow is a ubiquitous effect. The local orientation director $\nb$ and the order parameter $f$ are directly correlated with the principal direction and intensity of the shear rate~\cite{ref19,ref20}. Accordingly, scanning $\nb$ and $f$ should allow one to retrieve the time-dependent shear rate distribution, which is of particular interest when studying microfluidic and biofluidic systems~\cite{ref21,ref22,ref23,ref24}. The currently available particle imaging velocimetry (PIV) technique, which measures the flow velocity profile by tracking the displacements of fluorescent microspheres~\cite{ref25,ref26}, requires heavy accumulation and post-treatment of image frames. This limits access to the local real-time observation of dynamic systems. Moreover, the signal-to-noise ratio and the spatial resolution of PIV deteriorate when the principal interest is in shear (gradient of velocity). Our approach aims to achieve direct measurement and fast scanning of the local shear rate by instantly detecting the collective orientation of nanorods in a small focal volume. As a proof of concept, we demonstrate tomographic mapping of the shear distribution in a microfluidic channel using scanning confocal microscopy.

\section*{Polarized photoluminescence from assembled nanorods}
\begin{figure*}
\begin{center}
\includegraphics[width=.95\textwidth]{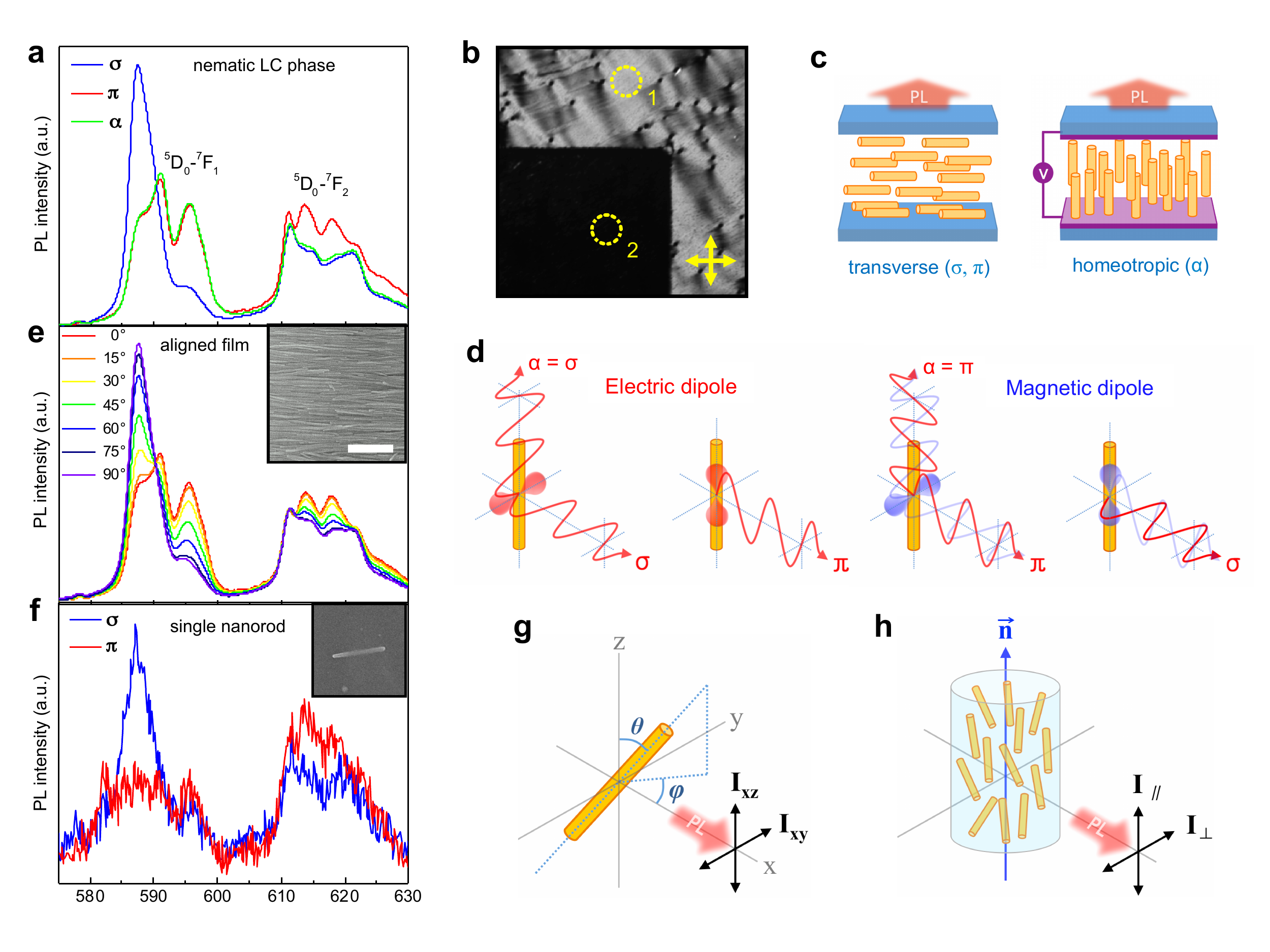}
\caption{\textbf{Polarized photoluminescence of LaPO$_4$:Eu nanorods.} (a) Photoluminescence (PL) spectra in the $\sigma$, $\pi$ and $\alpha$ configurations obtained from the nematic LC suspension of LaPO$_4$:Eu nanorods with transverse and homeotropic orientations. (b) Optical transmission microscopy image of the electro-optical switching cell filled with the nematic suspension. The image dimensions are 1.8 mm $\times$ 1.8 mm. The cell was placed between crossed polarizers (indicated by arrows). The inverted L-shaped bright region is where the nematic domain was transversely oriented due to the capillary invasion. The wavy texture was generated due to the 20-$\mu$m-thick silica bead spacers (black dots dispersed in the bright region). The dark rectangular region corresponds to the area (delimited by the top and bottom transparent electrodes of the cell) where a longitudinal electric field was applied to obtain homeotropic orientation. Dotted circles ‘1’ and ‘2’ indicate the transverse and homeotropic zones where the polarized photoluminescence spectra shown in a were collected. (c) Schematic illustration of the transverse and homeotropic orientations of the nanorods in the cell. (d) Schematic illustration of the contributions of electric dipole transition and magnetic dipole transition in the $\sigma$, $\pi$ and $\alpha$ configurations. (e) Polarized photoluminescence spectra obtained from an aligned thin film of LaPO$_4$:Eu nanorods fabricated by the shear-directed assembly. Analyser angle $\theta$ was changed in 15$^\circ$ steps. Inset: scanning electron microscopy (SEM) image of the film (scale bar, 200 nm). (f) Polarized photoluminescence spectra (σ and π configurations) obtained from a single nanorod using a confocal microscope. The position and orientation of the nanorod were detected using SEM and atomic force microscopy (AFM; Supplementary Fig. 2). Inset: SEM image of the same nanorod (rod length of $~$500 nm). (g) Schematic illustration of a nanorod with an angular orientation ($\theta$,$\phi$) with respect to the photoluminescence measurement plane ($y$–$z$ plane). (h) Schematic illustration of a preferential partial orientation of colloidal nanorods due to an external stimulus and of transverse photoluminescence measurements $I_\parallel$ and $I_\perp$}\label{fig1}
\end{center}
\end{figure*}

Figure~\ref{fig1}a presents polarized photoluminescence spectra from a nematic LC suspension of LaPO$_4$:Eu nanorods modulated in an electro-optical cell. The most intense $^5$D$_0$–$^7$F$_1$ (magnetic dipole) transition and the adjacent $^5$D$_0$–$^7$F$_2$ (electric dipole) transition spectra, both consisting of multiple sublevel peaks, were collected under excitation of the $^7$F$_0$–$^5$L$_6$ transition at 394 nm (the excitation spectrum is shown in Supplementary Fig.~3). An optical transmission microscopy image of the cell placed between crossed polarizers (Fig.~\ref{fig1}b) shows a bright region exhibiting in-plane birefringence induced by the transverse rod alignment, as schematized in Fig.~\ref{fig1}c, left. A uniform LC domain orientation was directed by shear strain applied during the capillary invasion of the viscous suspension into the 20-$\mu$m-thick gap of the cell (for the cell geometry see Supplementary Fig. 1). The blue and red curves in Fig.~\ref{fig1}a were obtained from this transverse region (dotted circle labelled ‘1’ in Fig.~\ref{fig1}b) with an analyser perpendicular and parallel to the domain orientation, respectively. Considering the hexagonal symmetry of LaPO$_4$:Eu, with the crystallographic $c$ axis parallel to the long axis of the rod~\cite{ref27}, these two spectra correspond to the mutually orthogonal polarization components referred to as $\sigma$ and $\pi$
configurations (Fig.~\ref{fig1}d). The difference in their line shapes is due to the independent polarization of each sublevel emission originating from the crystal-field splitting~\cite{ref9}. When the electric field was applied longitudinally in the cell gap (Fig.~\ref{fig1}c, right), the nematic domain, initially in a transverse state, was switched to a homeotropic state where the birefringence vanished completely (dark square region in Fig.~\ref{fig1}b) as a consequence of the rod alignment being along the field normal to the substrate plane. The α spectrum (green line in Fig.~\ref{fig1}a) corresponding to the axial propagation (Fig.~\ref{fig1}d) was obtained from this homeotropic region (dotted circle ‘2’ in Fig.~\ref{fig1}b). In contrast to the $\sigma$ and $\pi$ spectra, the $\alpha$ spectrum was unchanged when rotating the analyser, because the axial symmetry of the crystal produces isotropic polarization contributions in the $c$ plane.

The $\sigma$ and $\pi$ spectra were also confirmed with two different types of sample, where the rod orientation was directly observable by scanning electron microscopy (SEM). First, a thin film with transverse rod alignment (Fig.~\ref{fig1}e, inset) was prepared by directed assem- bly of a nematic gel suspension~\cite{ref28}. The polarized photoluminescence spectra observed from this solid film (Fig.~\ref{fig1}e) with analyser angle $\theta$ perpendicular ($\theta = 90^o$) and parallel ($\theta=0^o$) to the rod orientation are identical to the $\sigma$ and $\pi$ spectra obtained from the LC sample (Fig.~\ref{fig1}a). Moreover, the tendency of the line shape variation with $\theta$ supports that the polarization of the Eu$^{3+}$ emission is subject to the uniaxial symmetry of the crystalline LaPO$_4$ matrix. A peak deconvolution study of the $^5$D$_0$–$^7$F$_1$ transition shows that the sublevel peak intensity $I_\theta$ as a function of $\theta$ closely fits the trigonometric equation $I_\theta = I_\sigma\sin 2\theta + I_\pi\cos 2\theta$, where $I_\sigma$ and $I_\pi$ indicate the peak intensities in the $\sigma$ and $\pi$ spectra (Supplementary Figs.~4–5). Furthermore, the polarized emission spectra taken from a single nanorod (Fig.~\ref{fig1}f) show the same $\sigma$ and $\pi$ spectral line shapes, verifying that the observed polarization behaviour originates exclusively from the intrinsic crystal structure and not from the collective effect of the assembled structure.

\section*{Three-dimensiontal orientation analysis}
Note that the $\alpha$ spectrum is identical to the $\pi$ spectrum in the (magnetic dipole) band and the $\sigma$ spectrum in the $^5$D$_0$–$^7$F$_2$ (electric dipole) band (Fig.~\ref{fig1}a). This can be understood considering that the radiation electric field is parallel to the electric dipole and perpendicular to the magnetic dipole (Fig.~\ref{fig1}d). This peculiar aspect of polarization offers the opportunity to measure the three-dimensional rod orientation precisely. Defining the rod orientation in the laboratory frame by the polar and azimuthal angles $(\theta,\phi)$ (Fig.~\ref{fig1}g), it is possible to express the two measured polarized photoluminescence intensities $I_{xy}$ and $I_{xz}$ (the two indices refer to the axis of propagation and the axis parallel to the analyser, respectively) as functions of $\theta$ and $\phi$. In the case of the magnetic dipole transition,
\begin{align}
I_{xy}&=I_\sigma\cos^2\theta+I_\pi\sin^2\theta,\\
I_{xz}&=I_\sigma\sin^2\theta\sin^2\phi+I_\pi\left(\cos^2\theta\sin^2\phi+\cos^2\phi\right)
\end{align}
where $I_\sigma$ and $I_\pi$ indicate the relative intensities of the $\sigma$ and $\pi$ configurations. By solving these equations simultaneously with their equivalents for the electric dipole transition (Supplementary Eqs.~(4) and (5)), one can determine the set of $(\theta,\phi)$ without consideration of the absolute intensities. Errors that may occur with the overall signal fluctuation from any extrinsic parameter can be avoided in this ratiometric line shape analysis. When regarding an ensemble of nanorods that acquires a partial orientation toward a preferential direction (Fig.~\ref{fig1}h), the order parameter, defined by $f = \langle 3\cos ^2\theta-1\rangle/2$, can be deduced from the following equations (written here also for the magnetic dipole transition):
\begin{align}
I_\perp&=I_\sigma\langle\cos^2\theta\rangle+I_\pi(1-\langle\cos^2\theta\rangle),\\
2I_\parallel&=I_\sigma(1-\langle\cos^2\theta\rangle)+I_\pi(1+\langle\cos^2\theta\rangle)
\end{align}
where $I_\perp$ and $I_\parallel$ indicate the two polarized photoluminescence intensities perpendicular and parallel to $\nb$. A complete description for obtaining $\nb$ (similar to the way of obtaining $\theta$ and $\phi$ for a single nanorod) and then $I_\perp$ and $I_\parallel$ from a system with unknown $\nb$ is provided in Supplementary Section II.

\section*{Flow shear measurement}
This capability to analyse the collective rod orientation was used to probe the local arrangement of nanorods induced by the shear flow. The stress-optical law describes the direct correlation between the shear rate ($\dot\gamma$) and $f$, which is proportional to the induced flow birefringence ($\Delta n$)~\cite{ref29,ref30,ref31}. Measuring $\Delta n$ allows us to deduce the value of $\dot\gamma$ and the related rheological parameters of fluids~\cite{ref32}. The rheological properties of bulk fluids have often been studied in this way~\cite{ref33}. However, so far, the stress-optical method has not been applicable to local measurements because birefringence is an integrated signal throughout the whole light pathway across the medium. Polarized photoluminescence, in contrast, enables a microscopic focal volume in the middle of the medium to be addressed. Local stress-optical analysis and three-dimensional mapping can therefore be attempted when taking advantage of the high sensitivity and resolution provided by confocal microscopy.

\begin{figure*}
\begin{center}
\includegraphics[width=.95\textwidth]{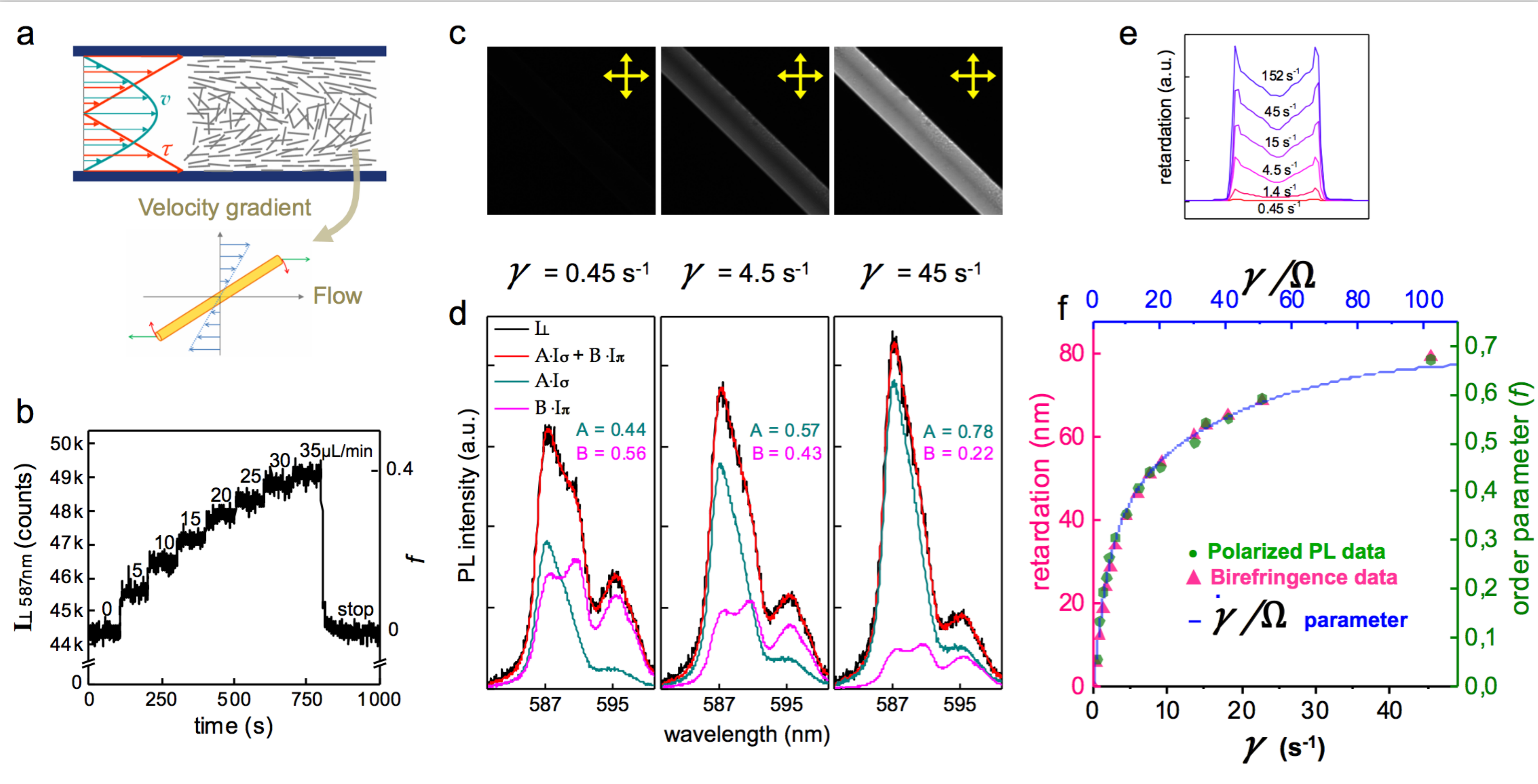}
\caption{ \textbf{Stress-optical measurements of a flowing nanorod suspension} (a) Schematic illustration of the orientation of colloidal nanorods under Poiseuille flow with typical velocity ($ν$) and shear stress ($\tau$) profiles. (b) Flow rate-dependent photoluminescence (PL) intensity at $\lambda$ = 587 nm from a colloidal LaPO$_4$:Eu nanorod suspension (0.9 vol\% in ethylene glycol) flowing in a capillary tube (diameter = 1 mm). Flow rate was increased in steps of 5 $\mu$L min$^{–1}$ each 100 s. (c) Optical transmission microscopy image of the same suspension flowing through a rectangular microfluidic channel (100 $\mu$m width, 50 $\mu$m height) for three different average shear rates ($\dot\gamma$). The channel was placed between crossed polarizers (indicated by arrows). (d) Polarized photoluminescence spectra of the $^5$D$_0$–$^7$F$_1$ transition (black line) and the fit (red line) by the weighted sum of Iσ (cyan line) and Iπ (pink line). Values of $A=\cos 2\theta$ and $B = 1 – \cos 2\theta$, the weight coefficients of $I_\sigma$ and $I_\pi$ obtained from the fit, are indicated. (e) Optical retardation ($\delta$) profiles across the channel width at different $\dot\gamma$. (f) Variations of average $\delta$ (pink triangles) and average order parameter obtained by analysis of the polarized photoluminescence spectra (green circles) as functions of $\dot\gamma$. The blue line is a theoretical calibration curve as a function of $\dot\gamma/\Omega$ fit with the stress-optical law.}\label{fig2}
\end{center}
\end{figure*}

We first studied the flow of a dilute colloidal nanorod suspension in a capillary tube. Figure~\ref{fig2}a schematically shows the general aspect of the rod orientation in a Poiseuille flow. Rods are highly oriented near the wall where a large shear stress ($\tau=\mu\dot\gamma$, where $\mu$ is the dynamic viscosity) is applied, and are disordered at the centre where $\tau$ approaches to zero. The photoluminescence spectrum observed from the whole capillary volume should be partially polarized in correlation with the average shear rate ($\dot\gamma$), which is proportional to the flow rate. Figure 2b shows the intensity variation of a polarized photoluminescence emission peak $I_\perp$ ($\lambda$ = 587 nm, observed with an analyser perpendicular to the capillary) while increasing the flow rate in steps. The initially disordered nanorods start to orient when the suspension flows. The increment of $I_\perp$ is directly proportional to $f$, according to Eq.~(3). Even a very small change in the flow rate of 5 $\mu$L min$^{–1}$ ($\sim 0.1$ mm s$^{–1}$ in average velocity) produces an appreciable change in $I_\perp$, ensuring the high sensitivity of the method.

To examine the validity of this polarized photoluminescence-based stress-optical measurement, it is necessary to compare it with the traditional method based on birefringence measurement. An experiment was implemented using a microfluidic channel with a rectangular cross-section, from which the photoluminescence and birefringence could be simultaneously observed. Figure~\ref{fig2}c displays the gradual increase of flow birefringence when increasing the flow rate, which can be converted into $\dot\gamma$. The corresponding polarized photoluminescence line shape, collected from a section of the channel volume with an analyser perpendicular to the flow direction, also changes towards the shape of the $\sigma$ spectrum, implying higher $f$ with higher $\dot\gamma$ (Fig.~\ref{fig2}d). The optical retardation ($\delta = \Delta n \times d$, where $d$ is the channel thickness) profiles across the channel width for different $\dot\gamma$ are shown in Fig.~\ref{fig2}e. These profiles represent the integrated birefringence through the channel depth. The $\delta$ value at the channel centre is non-zero due to the contribution of shear at the top and bottom surfaces of the channel. This illustrates why local stress-optical analysis cannot be achieved by birefringence measurements alone. In Fig.~\ref{fig2}f, the $\delta$ values averaged over the image plane are plotted (green circles) along with the $f$ values deduced from the line shape analysis of the polarized photoluminescence spectra (red triangles) collected over the same image plane. These two independently measured quantities display an identical evolution as a function of $\dot\gamma$ (bottom abscissa axis). An excellent agreement is also found with a theoretical calibration curve for $f$ (blue line) as a function of $\dot\gamma/\Omega$ (top abscissa axis)~\cite{ref34} when $\Omega$, the rotational diffusion coefficient, was set to be 0.5 s$^{−1}$ for the best fitting. Theoretically, $\Omega$ is given as
\begin{equation}
\Omega=\frac{3k_B T}{16\pi\eta_0a^3}\left(-1+2\log\frac{2a}{b}\right),
\end{equation}
where $k_B$ is the Boltzmann constant, $\eta_0$ is the solvent viscosity, and $a$ and $b$, respectively, are the half-length and equatorial radius of the particle~\cite{ref20}. The $\Omega$ value calculated with the measured viscosity and average nanorod size is 10 s$^{−1}$, which is an order of magnitude greater than 0.5 s$^{−1}$ from the calibration curve fitting. We estimate that the collective behaviour of nanorods with surface-charge-mediated long-range repulsive interactions or size polydispersity are responsible for such a difference.

\begin{figure*}
\begin{center}
\includegraphics[width=.95\textwidth]{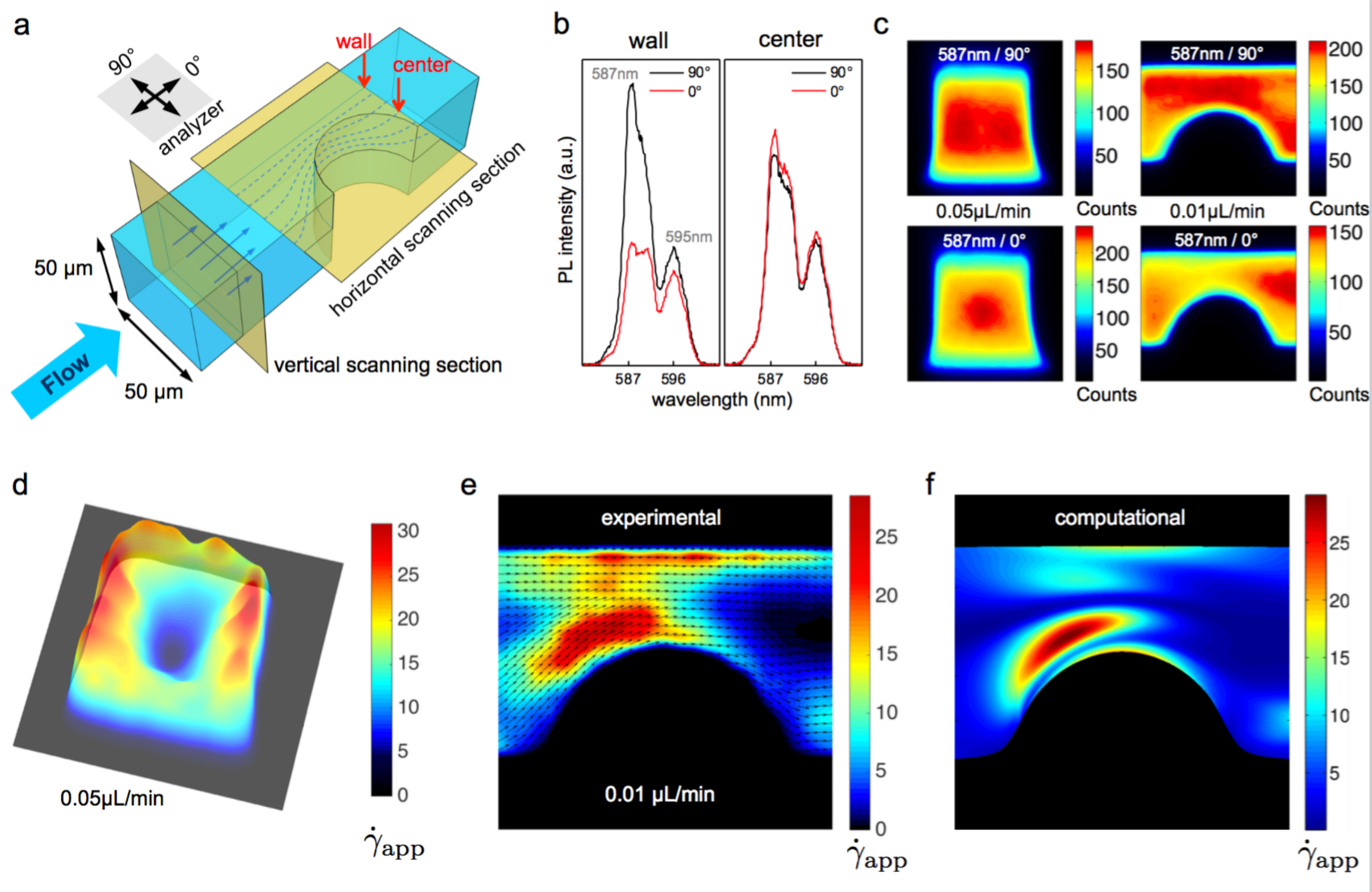}
\caption{\textbf{Tomographic stress-optical analysis of a microfluidic system.} (a) Schematic illustration of the microfluidic channel used for the local shear rate measurement. The channel has a square-shaped cross-section (50 $\times$ 50 $\mu$m) and a semicircular constriction (diameter = 50 $\mu$m). Two photoluminescence (PL) scanning sections are shown as yellow planes. The horizontal section is placed at the middle height of the channel and includes the constriction.
Analyser angles $\theta$ of $0^\circ$ and $90^\circ$ correspond to the directions parallel and perpendicular to the channel length on the observation plane parallel to the horizontal scanning section. (b) Polarized photoluminescence spectra ($\theta=0^\circ$ and $90^\circ$) of the $^5$D$_0$–$^7$F$_1$ transition collected at the channel wall and centre positions as indicated with red arrows in (a). (c) Colour maps of the polarized photoluminescence intensity ($\theta=0^\circ$ and $90^\circ$) centred at $\lambda$ = 587 nm selected by a bandpass filter. The left two images were obtained on the vertical scanning section with a flow rate of 0.05 $\mu$L min$^{–1}$. The right two images were obtained on the horizontal scanning section with a flow rate of 0.01 $\mu$L min$^{–1}$. The colour bars are scaled in number of counts per pixel. (d) Three-dimensional colour map of the apparent shear rate on the vertical scanning section. (e) Colour map of $\dot\gamma_\textrm{app}$ superposed with a vector map of $\nb$ on the horizontal scanning section. The maps in (d) and (e) were obtained by stress-optical analysis of the polarized photoluminescence intensity maps shown in (d). (f) Computationally obtained $\dot\gamma_\textrm{app}$ map with the Peclet number set to 5.}\label{fig3}
\end{center}
\end{figure*}
The above results guarantee a reliable stress-optical measurement by analysing the polarized photoluminescence of colloidal LaPO$_4$:Eu nanorods. On this basis, local detection of $\dot\gamma$ was tested for a geometrically complex flow generated in a microfluidic channel with a constriction (Fig.~\ref{fig3}a). Confocal microscopy with laser excitation at 394 nm ($^7$F$_0$–$^5$L$_6$ transition) was performed for the local photoluminescence measurement. The polarized photoluminescence spectra emitted from focal spots positioned at the channel wall and centre (indicated by red arrows in Fig.~\ref{fig3}a) are plotted in Fig.~\ref{fig3}b. At the wall, where $\dot\gamma$ reaches its maximum, the two spectra obtained with analyser angles of $0^o$ and $90^o$ exhibit contrasting line shapes that are close to the $\pi$ and $\sigma$ spectra, respectively. However, at the channel centre, the two spectra are almost identical, because $\dot\gamma$ approaches zero. When the flow rate was abruptly changed, the resulting spectral fluctuation at a fixed focal position could be recorded with the time resolution available with the spectrometer.
The scanning operation of the confocal microscope allowed the construction of tomographic maps of such a local shear measurement over any scanning section. Figure~\ref{fig3}c presents polarized photoluminescence intensity maps for a peak wavelength of 587 nm over vertical and horizontal scanning sections (Fig.~\ref{fig3}a). From each set of these intensity maps, a corresponding map of the apparent shear rate ($\dot\gamma_\textrm{app}$) can be constructed by simple image processing using Eqs.~(3) and (4) and the $f$ vs. $\dot\gamma$ calibration curve in Fig.~\ref{fig3}f. Note that, because the $f$ vs. $\dot\gamma$ curve is not linear, the accuracy of the determination of $\dot\gamma_\textrm{app}$ should depend on its value. However, the relative uncertainty, $\Delta\dot\gamma/\dot\gamma$, estimated from the derivative of the $f$ vs. $\dot\gamma$ curve, varies weakly as soon as $\dot\gamma/\Omega>1$ (Supplementary Fig. 6). This is advantageous for an accurate determination of $\dot\gamma_\textrm{app}$ in a wide range extending over 100 s$^{−1}$. The $\Delta\dot\gamma/\dot\gamma$ value in our experiment is smaller than 30\%. Figure~\ref{fig3}d shows a surface plot of the $\dot\gamma_\textrm{app}$ profile over the vertical scanning section. In this case, the local director of $\dot\gamma$, that is, the principal shear direction, is constantly normal to the scanning plane. This $\dot\gamma_\textrm{app}$ profile is similar to the inverted pyramidal shape theoretically predicted for a Poiseuille flow in a rectangular channel. This result implies the near-Newtonian rheology of the dilute LaPO$_4$:Eu nanorod suspension, which thus seems to be suited for performing the stress-optical analysis. Meanwhile, the $\dot\gamma_\textrm{app}$ map constructed over the horizontal scanning section with the semicircular constriction (Fig.~\ref{fig3}e, the colour represents the quantity of $\dot\gamma_\textrm{app}$ and the arrows indicate $\nb$) shows an unexpected asymmetric profile. Considering the small channel dimension and flow rate (Reynolds number of $\sim 10^{-3}$), inertia plays a negligible role here and the flow regime is purely viscous, which should lead to a fore-aft symmetry for Newtonian fluids~\cite{ref35}, as can be seen from a theoretically modelled $\dot\gamma$ map (Supplementary Fig. 8a). However, in Fig.~\ref{fig3}e, the region of maximum $\dot\gamma_\textrm{app}$ (that is, maximum $f$) is deviated towards the upstream (left side) and away from the constriction wall. Similar asymmetric birefringence profiles have also been reported in the study of viscoelastic polymer suspensions~\cite{ref36}.

\section*{Computational analysis}
Examining the asymmetry of the $\dot\gamma_\textrm{app}$ map (Fig.~\ref{fig3}e) is imperative for establishing a reliable stress-optical analysis. Therefore, we con- ducted a computational analysis of the rod-orientation dynamics. The details of the method and results are provided in Supplementary Section III and Supplementary Fig.~7. This study reveals that the discrepancy between $\dot\gamma_\textrm{app}$ and the real shear rate $\dot\gamma$ originates in the advection and non-instantaneous reorientation of the nanorods in a non-homogeneous flow. When the streamlines are not parallel around the constriction, the nanorods advected by the flow experience time-dependent local shear that varies over a typical timescale $\tau_a= H/U$ (where $H$ is the channel width and $U$ the flow velocity) taken to flow over the constriction. However, the response of the probability distribution for the nanorods’ orientation occurs on the rotational diffusion timescale $\tau_d = Ω^{–1}$. The Peclet number is the ratio of these two timescales, $\mbox{Pe} = U/H\Omega$, and measures the relative rates of advection and diffusion. For small $\mbox{Pe}$, the response is instantaneous, and the stress-optical law holds everywhere, whereas for intermediate to large $\mbox{Pe}$, the orientation of individual particles depends on the local shear but also on the shear to which the particles have been exposed previously, leading to history effects. Indeed, the computationally-obtained $\dot\gamma_\textrm{app}$ map for $\mbox{Pe} = 0.5$ (Supplementary Fig.~8b) is almost identical to the theoretical $\dot\gamma$ map, which is symmetric (Supplementary Fig. 8a). However, the $\mbox{Pe}$ given for our experiment is 5 when applying the $\Omega = 0.5 $s$^{–1}$ obtained from the calibration curve (Fig.~\ref{fig2}f). History effects are thus non-negligible and explain the discrepancy between the apparent and real shear rates: the computationally obtained $\dot\gamma_\textrm{app}$ map for $\mbox{Pe} = 5$ (Fig.~\ref{fig3}f) matches well the experimental $\dot\gamma_\textrm{app}$ map (Fig.~\ref{fig3}e), both qualitatively (highly ordered upstream and rapid loss of orientation coherence downstream) and quantitatively.

These computational results suggest that, to produce reliable $\dot\gamma_\textrm{app}$ maps, the $\mbox{Pe}$ number needs to be small enough to suppress the history effects. This can be realized by reducing the particle size, which would rapidly increase $\Omega$ (Equation~(5)). We estimate that if the LaPO$_4$:Eu nanorod size is reduced by a factor of three ($\mbox{Pe} = 0.2$), the stress-optical analysis and the tomographic $\dot\gamma$ mapping would be satisfactory for most microfluidic systems of interest.

\section*{Conclusions}
We have presented a simple method to measure the polarized photoluminescence spectra (in $\sigma$, $\pi$ and $\alpha$ configurations) of LaPO$_4$:Eu nanorods from their electrically modulated liquid-crystalline phase. The three-dimensional orientation of an individual nanorod or the director ($\nb$) and order parameter ($f$) of a rod assembly can be precisely determined by analysing their polarized photoluminescence line shape. This approach allowed us to investigate the rod-orientation dynamics of the colloidal nanorods flowing in a microfluidic channel. The local shear rate ($\dot\gamma$) profiles over sections of the fluid volume were deduced based on the stress-optical law. A reliable estimation of the $\dot\gamma$ profile was obtained over a region where the streamlines are parallel. However, a discrepancy between theory and experiment was found for non-parallel flows involving rod advection and reorientation. A computational study verified that this discrepancy is due to the non-instantaneous reorientation of nanorods and that it could be effectively suppressed by decreasing the nanorod size. With further optimization of the nanorod size, this technique is promising in that it yields a straightforward stress-optical method for tomographic mapping and real-time monitoring of local $\dot\gamma$ with the high spatial resolution necessary for applications in microfluidics and biofluidics. Furthermore, the presented orientation analysis might be exploited for the study of the complex dynamics of other microscopic systems (such as cells, genes and enzymes) by using rare-earth luminophores as orientation markers.

\section*{Methods}
Methods and any associated references are available in the online version of the paper.

\begin{acknowledgements}
The authors thank C. Frot and N. Taccoen for the fabrication of microfluidic channels, C. Henry de Villeneuve for atomic force microscopy and A. Agrawal for graphics. This research was partially supported by LASERLAB-EUROPE (grant agreement no. 284464 from the European Community’s Seventh Framework Programme). G.A., E.F. and C.N.B. acknowledge funding by the ERC under grant agreement 278248 (Multicell).
\end{acknowledgements}


\end{document}